\documentclass{article}
\usepackage{spconf,amsmath,graphicx,amssymb}
\usepackage{color,multirow}
\usepackage{cite}
\usepackage{type1cm}
\usepackage{algorithm}
\usepackage{algcompatible}
\algblockdefx{FORP}{ENDFORP}[1]%
  {\textbf{for}#1 \textbf{do in parallel}}%
  {\textbf{end for}}
\algblockdefx{Repeat}{Until}[1]%
  {\textbf{repeat}#1}%
  {\textbf{until} converged}
\algblockdefx{If}{Else}[1]%
  {\textbf{if} #1}%
  {\textbf{else}}
  
\usepackage{setspace}   
\usepackage{url}

\newcommand{\mysubsection}[1]{\vspace{-11pt}\subsection{#1}}

\title{Hierarchical Diffusion Models \\ for Singing Voice Neural Vocoder}
\name{Naoya Takahashi, Mayank Kumar, Singh, Yuki Mitsufuji}
\address{Sony Group Corporation, Japan}
\begin{document}
\ninept


\maketitle
\begin{abstract}
Recent progress in deep generative models has improved the quality of neural vocoders in speech domain. However, generating a high-quality singing voice remains challenging due to a wider variety of musical expressions in pitch, loudness, and pronunciations. In this work, we propose a hierarchical diffusion model for singing voice neural vocoders. The proposed method consists of multiple diffusion models operating in different sampling rates; the model at the lowest sampling rate focuses on generating accurate low-frequency components such as pitch, and other models progressively generate the waveform at higher sampling rates on the basis of the data at the lower sampling rate and acoustic features. Experimental results show that the proposed method produces high-quality singing voices for multiple singers, outperforming state-of-the-art neural vocoders with a similar range of computational costs.
\end{abstract}
\section{Introduction}
\renewcommand{\thefootnote}{\fnsymbol{footnote}}
\label{sec:intro}
Neural vocoders generate a waveform from acoustic features using neural networks \cite{Aaron2016WN, Mehri2017SampleRNN, Prenger2019WaveGlow, Ping20WaveFlow, Yamamoto20PWG, Kong21DiffWave} and have become essential components for many speech processing tasks such as text-to-speech\cite{Aaron2016WN,Shen2018TTS,Li2019TTS}, voice conversion\cite{Qian19AutoVC, Li21StarGANv2VC,Agarwal2022SymNet}, and speech enhancement\cite{Maiti2019,Su2019HiFiGAN,liu2021voicefixer}, as they often operate in acoustic feature domains for the efficient modeling of speech signals. A number of generative models have been adopted to neural vocoders such as autoregressive models \cite{Aaron2016WN,Mehri2017SampleRNN,Kalchbrenner2018WaveRNN}, generative adversarial networks (GANs) \cite{Donahue2019WaveGAN,Kumar2019MelGAN,Yamamoto20PWG,Kong2020HiFiGAN}, and flow-based models \cite{Prenger2019WaveGlow, Ping20WaveFlow}.

Recently, diffusion models \cite{Song2019NCSN, Ho2020DDPM} have attracted increasing attention in a wide range of areas \cite{Lu2022cDPM,Lai2022} as they are shown to generate high-fidelity samples.
Denoising diffusion probabilistic models (DDPMs) \cite{Ho2020DDPM} gradually convert a simple distribution such as an isotropic Gaussian into a complicated data distribution using a Markov chain.
To learn the conversion model, DDPMs use another Markov chain called a \textit{forward process} (or \textit{diffusion process}) to gradually convert the data to the simple target prior (e.g. isotropic Gaussian) by adding noise. Since the forward process is done without any trainable model, DDPMs avoid the challenging “posterior collapse” issues caused by the joint training of two networks such as a generator and discriminator in GANs \cite{Goodfellow2014GAN} or an encoder and decoder in variational autoencoder (VAE)\cite{Kingma2014VAE}. 
Although the data likelihood is intractable, diffusion models can be efficiently trained by maximizing the evidence lower bound (ELBO).

Diffusion models have been adopted to neural vocoders \cite{Kong21DiffWave,Chen2021WaveGrad}. Although they are shown to produce high-quality speech data, the inference speed is relatively slow compared with other non-autoregressive model-based vocoders as they require many iterations to generate the data. PriorGrad \cite{Lee22PriorGrad} addresses this problem by introducing a data dependent prior, specifically, Gaussian distribution with a diagonal covariance matrix whose entries are frame-wise energies of the mel-spectrogram. As the noise drawn from the data dependent prior is closer to the target waveform than the noise from standard Gaussian, PriorGrad achieves faster convergence
and inference with superior performance. Koiszumi et al. \cite{Koizumi22SpecGrad} further improve the prior by incorporating the spectral envelope of the mel-spectrogram to introduce the noise that is more similar to the target signal. 

However, many existing neural vocoders focus on speech signals. We found state-of-the-art neural vocoders provide insufficient quality when they are applied to a singing voice, possibly due to the scarcity of large-scale clean singing voice datasets and the wider variety in pitch, loudness, and pronunciations owing to musical expressions, which is more challenging to model. To overcome this problem, we propose a hierarchical diffusion model that learns multiple diffusion models at different sampling rates. 
The diffusion models are conditioned on acoustic features and the data at the lower sampling rate, and can be trained in parallel. During the inference, the models progressively generate the data from the low to high sampling rate.
The diffusion model at the lowest sampling rate focuses on generating low frequency components, which enables accurate pitch recovery, while those at higher sampling rates focus more on high frequency details. This enables powerful modeling capability of a singing voice. In our experiment,  we apply the proposed method to PriorGrad and show that the proposed model generates high-quality singing voices for multiple singers, outperforming the state-of-the-art PriorGrad and Parallel WaveGAN\cite{Yamamoto20PWG} vocoders.

In image generation tasks, cascading generative models are shown to be effective \cite{Menick2019, Razavi2019VQVAE2, ho2021cascaded}. However, to our knowledge, this is the first work that uses multiple diffusion models in different sampling rates for progressive neural vocoding.
Our contributions are;
(i) we propose a hierarchical diffusion model-based neural vocoder to generate a high-quality singing voice, and
(ii) we apply the proposed model to PriorGrad and show in our experiments that the proposed method outperforms state-of-the-art neural vocoders, namely, PriorGrad and Paralell WaveGAN with a similar computational cost.
Audio samples are available at our website\footnote{\url{https://t-naoya.github.io/hdm/}\label{fn:demo}}.

\begin{figure*}[t]
  \centering
  \includegraphics[width=0.75\linewidth]{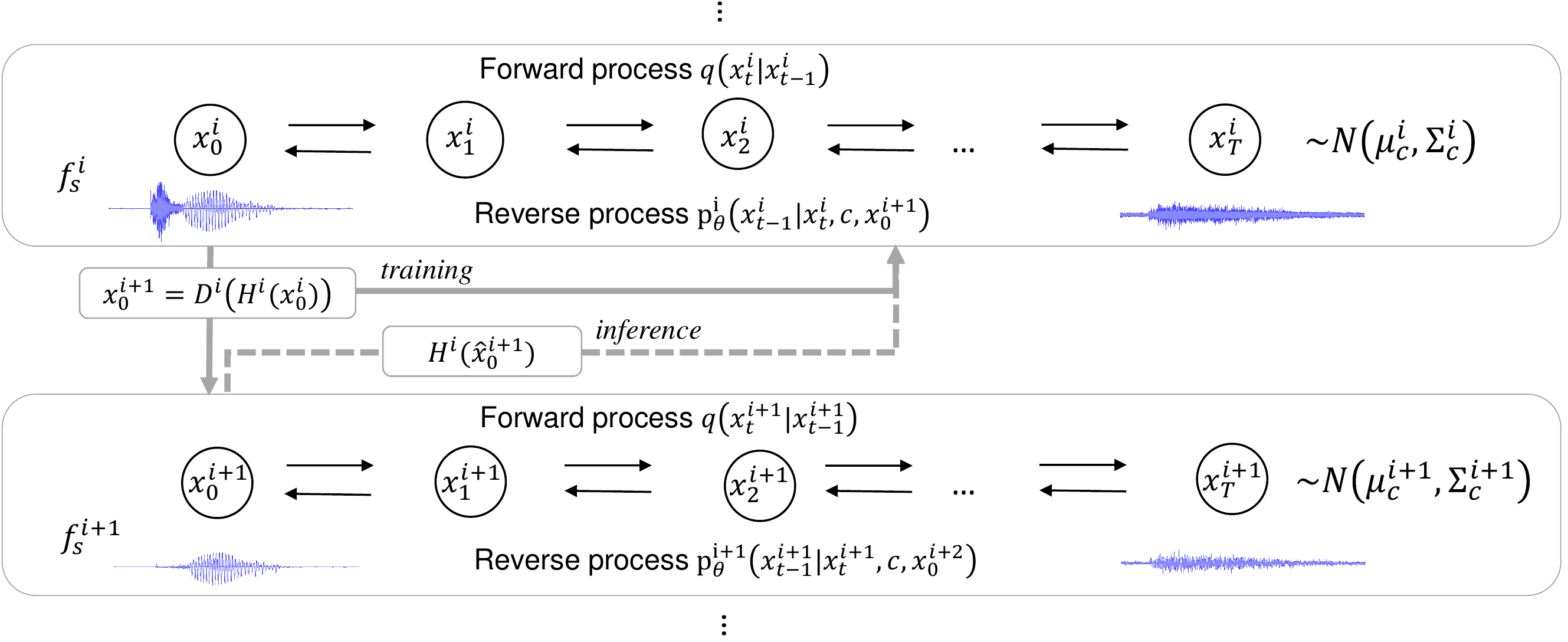}
  \caption{Overview of proposed hierarchical diffusion model combined with PriorGrad \cite{Lee22PriorGrad}. Diffusion models are trained at multiple sampling rates $f_s^1>\cdots>f_s^N$ independently. Each diffusion model is conditioned on acoustic features $c$ and data at the lower sampling rate $x_0^{i+1}$. During inference, the anti-aliasing filter $H^i$ is applied to the data generated at the lower sampling rate and used for conditioning.}
  \label{fig:overview}
\end{figure*}

\section{Prior work}
\subsection{Denoising diffusion probabilistic models (DDPM)}
\label{sec:osvc}
DDPMs are defined by two Markov chains, the \textit{forward} and \textit{reverse} processes. The \textit{forward process} gradually diffuses the data $x_0$ into a standard Gaussian $x_T$ as:
\begin{equation}
    q(x_{1:T}|x_0) =\prod^T_{t=1} q(x_t|x_{t-1}),
    \label{eq:fw}
\end{equation}
where $q(x_t|x_{t-1}):= N(x_t;\sqrt{1-\beta_t}x_{t-1}, \beta_t \textbf{I})$ is a transition probability at a time-step $t$ that adds small Gaussian noise on the basis of a noise schedule $\beta_t\in\{\beta_1,\cdots,\beta_T\}$. This formulation enables to directly sample $x_t$ from $x_0$ as: 
\begin{equation}
x_t=\sqrt{\bar{\alpha}_t}x_0 + \sqrt{(1-\bar{\alpha}_t)}\epsilon,
    \label{eq:xt}
\end{equation}
where $\alpha_t=1-\beta_t$, $\bar{\alpha}_t=\prod_{s=1}^t\alpha_s$, and $\epsilon\sim\mathcal{N}(\textbf{0},\textbf{I})$.
The \textit{reverse process} gradually transforms the piror noise $p(x_T)=\mathcal{N}(x_T;\textbf{0},\textbf{I})$ to data as:
\begin{equation}
    p(x_{0:T}) =p(x_T)\prod^T_{t=1} p_\theta(x_{t-1}|x_t),
    \label{eq:rev}
\end{equation}
where $p_\theta(x_{t-1}|x_t):=\mathcal{N}(x_{t-1};\mu_\theta(x_t,t),\sigma_\theta^2(x_t,t)\textbf{I})$ is a transition probability that corresponds to the reverse of $q(x_t|x_{t-1})$ and is modeled by a deep neural network parameterized by $\theta$.
Ho et al.\cite{Ho2020DDPM} show that $p_\theta(x_{t-1}|x_t)$ can be given by
\begin{eqnarray}
    \mu_\theta(x_t,t)=\frac{1}{\sqrt{\alpha_t}}(x_t-\frac{\beta_t}{\sqrt{1-\bar{\alpha}_t}}\epsilon_\theta(x_t,t)),
\end{eqnarray}
and $\sigma_\theta^2(x_t,t)=\frac{1-\bar{\alpha}_{t-1}}{1-\bar{\alpha}_t}\beta_t$, where $\epsilon_\theta(x_t,t)$ is a deep neural network that predics the noise $\epsilon$ added at time $t$ in \eqref{eq:xt}. The model $\epsilon_\theta(x_t,t)$ can be optimized by maximizing the ELBO:
\begin{equation}
    ELBO =C-\sum_{t=1}^T\kappa_t\mathbb{E}_{x_0,\epsilon}[||\epsilon-\epsilon_\theta(x_t,t)||^2],
    \label{eq:elbo}
\end{equation}
where C is a constant, $\kappa_t=\frac{\beta_t}{2\alpha(1-\bar{\alpha}_{t-1})}$ for $t>1$ and $\frac{1}{2\alpha}$ for $t=1$. As suggested in \cite{Ho2020DDPM}, many followup works instead use a simplified loss function by setting $\kappa_t=1$ \cite{Ho2020DDPM,Kong21DiffWave,Lee22PriorGrad}. DDPM have been adopted to neural vocoders by conditioning the noise estimation network on acoustic features $c$ as $\epsilon_\theta(x_t,c,t)$ \cite{Chen2021WaveGrad,Kong21DiffWave}. Starting from the noise sampled from the prior $x_T$, the DDPM-based vocoders iteratively denoise the signal $x_t$ on the basis of the condition $c$ to obtain the corresponding waveform $x_0$.

\subsection{PriorGrad}
Although the standard Gaussian prior in DDPMs provides a simple solution without any assumption on the target data, it requires many steps to obtain high-quality data, which hinder efficient training and sampling. To improve the efficiency in the neural vocoder case, PriorGrad \cite{Lee22PriorGrad} uses an adaptive prior $\mathcal{N}(\textbf{0},\mathbf{\Sigma_c})$, where the diagonal variance $\mathbf{\Sigma_c}$ is computed from a mel-spectrogram $c$ as $\mathbf{\Sigma_c}= diag[(\sigma_0^2, \cdots, \sigma_L^2)]$ and $\sigma_i^2$ is a normalized frame-level energy of the mel-spectrogram at the $i$th sample. The loss function is modified accordingly to 
\begin{equation}
L=\mathbb{E}_{x_0,\epsilon,t}[||\epsilon-\epsilon_\theta(x_t,c,t)||^2_{\Sigma^{-1}}],
\end{equation}
where $||\mathbf{x}||^2_{\Sigma^{-1}}=\mathbf{x}^\top\Sigma^{-1}\mathbf{x}$. Intuitively, as the power envelope of the adaptive prior is closer to that of the target signal than that of the standard Gaussian prior, the diffusion model can require fewer time steps to converge and be more efficient.

\section{Proposed method}
\subsection{Hierarchical diffusion probabilistic model}
\label{sec:HDPM}
Although PriorGrad shows promising results on speech data, we found that the quality is unsatisfactory when it is applied to a singing voice, possibly due to the wider variety in pitch, loudness, and musical expressions such as vibrato and falsetto. To tackle this problem, we propose to improve the diffusion model-based neural vocoders by modeling the singing voice in multiple resolutions. An overview is illustrated in Figure \ref{fig:overview}. Given multiple sampling rates $f_s^1>f_s^2>\cdots>f_s^N$, the proposed method learns diffusion models at each sampling rate independently. 
The reverse processes at each sampling rate $f_s^{i}$ are conditioned on common acoustic features $c$ and the data at the lower sampling rate $f_s^{i+1}$ as $p^i_\theta(x^i_{t-1}|x^i_t, c, x^{i+1}_0)$ except the model at the lowest sampling rate, which is conditioned only on $c$.
During the training, we use the ground truth data $x_0^{i+1}=D^i(H^i(x^i_0))$ to condition the noise estimation models $\epsilon^i_\theta(x^i_t, c, x^{i+1}_0,t)$, where $H^i(.)$ denotes the anti-aliasing filter and $D^i(.)$ denotes the downsampling function for the signal at the sampling rate of $f_s^i$. 
\begin{algorithm}
\caption{Training of Hierarchical PriorGrad}\label{alg:train}
\begin{algorithmic}
\State \textbf{Given:} $f_s^1>\cdots>f_s^N$
\FORP { $i=1,\cdots,N$}
\Repeat
\State $x^i_0 \sim q^i_{data}, \epsilon^i\sim\mathcal{N}(\textbf{0},\mathbf{\Sigma_c}),t\sim\mathcal{U}([0,\cdots,T])$
\State $x_0^{i+1}=D^i(H^i(x^i_0))$ if $i<N$; else $x_0^{i+1}=$~\textit{Null}
\State $x^i_t=\sqrt{\bar{\alpha}_t}x^i_0 + \sqrt{(1-\bar{\alpha}_t)}\epsilon$
\State $L=||\epsilon^i - \epsilon_\theta(x^i_t, c, x^{i+1}_0,t)||^2$
\State Update the model parameter $\theta$ with $\nabla_\theta L$
\Until
\ENDFORP
\end{algorithmic}
\end{algorithm}
Since the noise $\epsilon$ is linearly added to the original data $x_0$ as in \eqref{eq:xt} and the model has direct access to the ground truth lower-sampling rate data $x^{i+1}_0$, the model can more simply predict the noise for low-frequency components from $x^i_t$ and $x^{i+1}_0$ by avoiding the complicated acoustic feature-to-waveform transformation. This enables the model to focus more on the transformation of high-frequency components. 
At the lowest sampling rate $f_s^N$ (we use 6 kHz in our experiments), the data $x_0^N$ become much simpler than that at the original sampling rate, and the model can focus on generating low-frequency components, which is important for accurate pitch recovery of a singing voice.

\begin{figure}[t]
  \centering
  \includegraphics[width=0.7\linewidth]{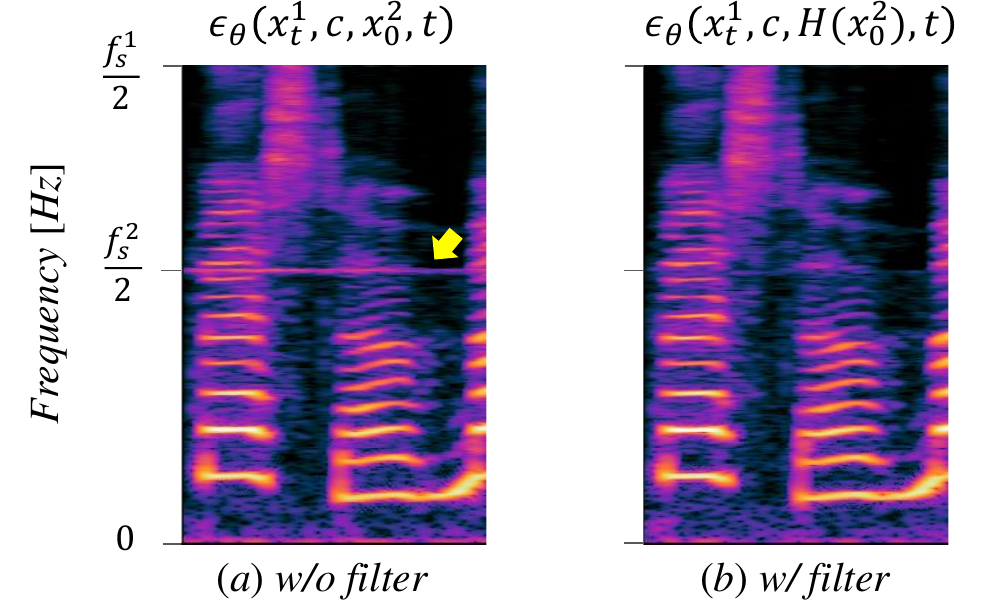}
  \caption{Anti-aliasing filter effect in the case of $N=2, f_s^1=24000, f_s^2=6000$.}
  \label{fig:filter}
\end{figure}

During inference, we start by generating the data at the lowest sampling rate $\hat{x}_0^N$ and progressively generate the data at the higher sampling rate $\hat{x}_0^i$ by using the generated sample $\hat{x}_0^{i+1}$ as the condition. 
In practice, we found that directly using $\hat{x}_0^{i+1}$ as the condition often produces noise around the Nyquist frequencies of each sampling rate, $\frac{f_s^2}{2},\cdots,\frac{f_s^N}{2}$, as shown in Figure \ref{fig:filter}~(a). 
This is due to the gap between the training and inference mode;  the ground truth data used for training $x_0^{i+1}=D^i(H^i(x^i_0))$ do not contain a signal around the Nyquist frequency owing to the anti-aliasing filter and the model can learn to directly use the signal upto the Nyquist frequency, while the generated sample used for inference $\hat{x}_0^{i+1}$ may contain some signal around there due to the imperfect predictions and contaminate the prediction at a higher sampling rate.
To address this problem, we propose to apply the anti-aliasing filter to the generated lower-sampling-rate signal to condition the noise prediction model as
\begin{equation}
\hat{\epsilon} = \epsilon^i_\theta(x^i_t, c, H(\hat{x}^{i+1}_0),t).
\label{eq:infer}
\end{equation}
As shown in Figure \ref{fig:filter}~(b), this removes the noise around the Nyquist frequencies and improves the quality. We summarize the training and inference procedures for the combination with PriorGrad in Algorithms \ref{alg:train} and \ref{alg:infer}, respectively.

The proposed hierarchical diffusion model can be combined with many types of diffusion models such as DiffWave \cite{Kong21DiffWave}, PriorGrad \cite{Lee22PriorGrad} and SpecGrad \cite{Koizumi22SpecGrad}.

\begin{algorithm}
\caption{Inference of Hierarchical PriorGrad}\label{alg:infer}
\begin{algorithmic}
\State \textbf{Given:} $f_s^1>\cdots>f_s^N$
\State $\hat{x}_0^{N+1}=$~\textit{Null}
\FOR { $i=N,N-1,\cdots,1$}
\State $x^i_T \sim\mathcal{N}(\textbf{0},\mathbf{\Sigma_c})$
\FOR { $t=T,T-1,\cdots,1$}
\State $\mathbf{z}\sim\mathcal{N}(\textbf{0},\mathbf{\Sigma_c})$ if $T>1$; else $\mathbf{z}=0$
\State $x^i_{t-1} = \frac{1}{\sqrt{\alpha_t}}(x^i_t-\frac{\beta_t}{\sqrt{1-\bar{\alpha}_t}}\epsilon_\theta(x^i_t,c,H(\hat{x}^{i+1}_0),t))+\sigma_t\mathbf{z}$
\ENDFOR
\ENDFOR ~~$x_0=x_0^1$
\end{algorithmic}
\end{algorithm}

\subsection{Network architecture}
As in PriorGrad \cite{Lee22PriorGrad}, we also base our model architecture on DiffWave \cite{Kong21DiffWave}. The network consists of $L$ residual layers with bidirectional dilated convolution and repeated dilation factors. The layers are grouped into $m$ blocks, and each block consists of $l=\frac{L}{m}$ layers with dilation factors of $[1,2,\cdots,2^{l-1}]$ (Please refer to \cite{Kong21DiffWave} for more details). DiffWave and PriorGrad use $L=30, l=10$ to cover large receptive fields. Our approach can take advantage of modeling in the lower sampling rates and a smaller network can cover the long signal length in seconds. Thus, we reduce the size to $L=24, l=8$ for all models in different sampling rates to mitigate the increase of computational cost due to the multi-resolution modeling. As shown in our experiments, this hyperparameter provides nearly the same computational cost as the original diffusion model when $N=2$. 

\begin{figure}[t]
  \centering
  \includegraphics[width=0.8\linewidth]{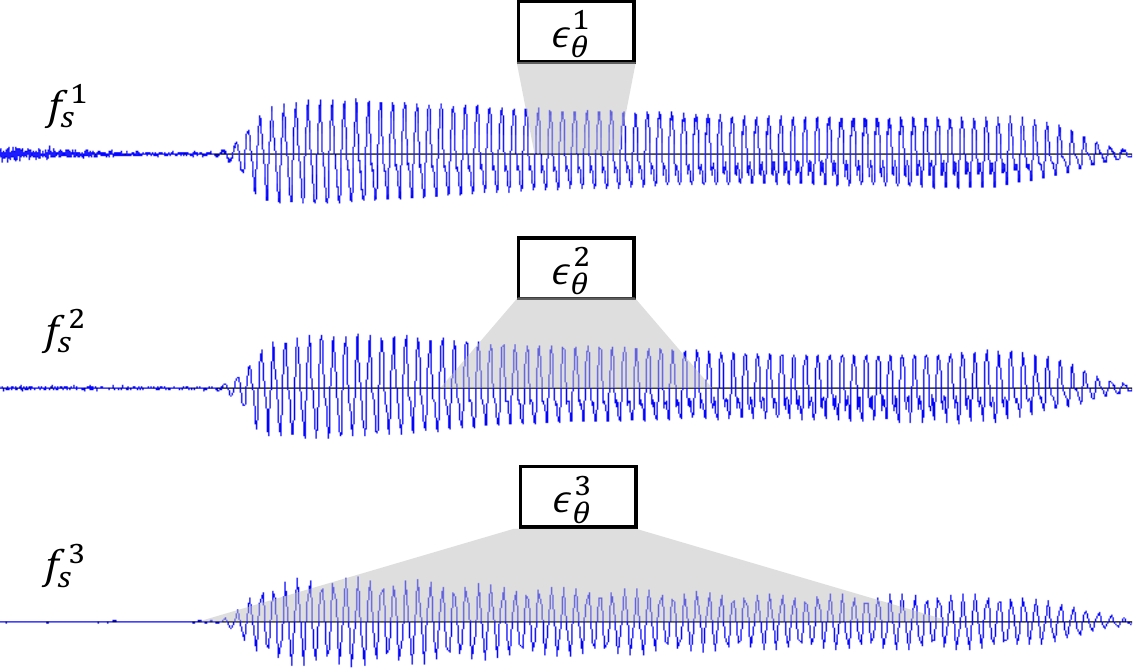}
  \caption{Receptive field at different sampling rates. The same architecture covers a longer time period at lower sampling rates.}
  \label{fig:rf}
\end{figure}

Using the same network architecture for all sampling rates effectively changes the receptive field of the models depending on the sampling rate, as illustrated in Figure \ref{fig:rf}. At the lower sampling rate, the model covers a longer time period and focuses on low-frequency components while it covers a shorter time period and focuses on high-frequency components at the higher sampling rate. This design matches our intention of the hierarchical diffusion model because the models are expected to directly use the conditioned data at the lower sampling rate $x_0^{i+1}$ upto the Nyquist frequency $\frac{f_s^{i+1}}{2}$ and focus on transforming the acoustic features to the waveform at the high-frequency.

\section{Experiments}
\mysubsection{Setup}
\textbf{Dataset}\hspace{2mm}
We use three singing voice datasets, NUS48E\cite{Duan13}, NHSS\cite{Sharma21NHSS}, and an internal corpus. NUS48E consists of twelve singers (six male and six female), four songs for each singer, while NHSS consists of ten singers (five male and five female), ten songs each. The internal corpus consists of eight singers (five male and three female), from 50 to 80 songs each. All data are resampled to 24k Hz.
We randomly split each dataset into 90\%--10\% for training and test sets. 
\\
\textbf{Hyper parameters}\hspace{2mm}
We follow the settings used in \cite{Kong21DiffWave, Lee22PriorGrad}; the models are trained for 1M steps with a batch size of 16, learning rate of $2\times10^{-4}$, and Adam optimizer \cite{Kingma2015Adam}. We use an 80-band mel spectrogram at log-scale as the acoustic feature. We set the FFT size to 2048 and hop size to 300. We adopt the proposed hierarchical diffusion model to PriorGrad \cite{Lee22PriorGrad} and evaluate two models, the 2-stage Hierarchical PriorGrad (HPG-2) with $(f_s^1,f_s^2) = (24k, 6k)$ and the 3-stage Hierarchical PriorGrad (HPG-3) with $(f_s^1,f_s^2,f_s^2) = (24k,12k,6k)$. The inference noise schedule is set to [0.0001, 0.001, 0.01, 0.05, 0.2, 0.5] with $T_{infer}=6$ as in \cite{Lee22PriorGrad}. 
\vspace{1mm}\\
\textbf{Baselines}\hspace{2mm}
We consider PriorGrad\footnote{\url{https://github.com/microsoft/NeuralSpeech}\label{fn:pg}} \cite{Lee22PriorGrad} and Parallel WaveGAN\footnote{\url{https://github.com/kan-bayashi/ParallelWaveGAN}\label{fn:PWG}} (PWG)\cite{Yamamoto20PWG} as baselines. 
All models are trained on the same singing voice dataset by following the instruction in the publicly available implementations.$^{\ref{fn:pg}\ref{fn:PWG}}$
\vspace{1mm}\\
\textbf{Evaluation metrics}\hspace{2mm}
For subjective evaluation, we rate naturalness of samples using five-point scale (1: Bad, 2: Poor, 3: Fair, 4: Good, 5: Excellent) and report the mean opinion score (MOS). We generate two 5-second singing voices for each of the 22 singers for each model and present to 20 raters in random orders. Raters evaluate the samples using headphones.

For the objective evaluation, we use five metrics; (i) real time factor (RTF), measured on a machine with a GeForce RTX 3090 to evaluate the computational cost, (ii) pitch mean absolute error (PMAE) between the ground truth and generated samples, where the pitch is extracted using the WORLD vocoder\cite{WORLD} (iii) voicing decision error (VDE) \cite{Nakatani2008}, which measures the portion of frames with voicing decision error, (iv) multi-resolution STFT error (MR-STFT) \cite{Yamamoto20PWG}, and (v) Mel cepstral distortion (MCD)\cite{Kubichek1993}. We evaluate the metrics on the test set and report the average values.

\begin{table}[t]
    \caption{\label{tab:subj} {\it MOS results with 95\% confidence interval.}}
    \vspace{2mm}
    \centering{
    \begin{tabular}{c  c } 
    \hline
    \textbf{Model}	&\textbf{MOS}\\
    \hline
    Ground Truth    & 4.66 $\pm$ 0.09\\
    \hline
    PWG \cite{Yamamoto20PWG}	&2.15 $\pm$ 0.13\\
    PriorGrad \cite{Lee22PriorGrad}	&3.60 $\pm$ 0.12\\
    \hline
    HPG-2 (Ours)	&\textbf{3.95} $\pm$ 0.13\\
    \hline
    \end{tabular}
    }
\end{table}

\mysubsection{Results}
We compare the MOS of the models that have similar RTFs  in Table \ref{tab:subj} (RTF values are shown in Table \ref{tab:obj}).  Unlike in the speech domain, Parallel WaveGAN (PWG) often suffers from unnatural shakes in pitch when the original singing voice has vibrato, which is possibly one of the reasons of the low MOS. PriorGrad does not produce such unnatural pitch shakes and obtains a higher MOS than PWG. The proposed HPG clearly outperforms the baselines, providing the best quality. 
We also evaluate a preference score, where raters are asked which of A and B is more natural, with A and B randomly chosen from HPG-2 and PriorGrad. We observe that 85.3\% of the time, raters prefer HPG-2.
Objective results are shown in Table \ref{tab:obj}. PMAE and VDE values of PWG are higher than other methods, which is consistent with the subjective observation. In contrast, PWG obtains lower MR-STFT and MCD values, which is inconsistent with the subjective results. This may be because PWG includes MR-STFT loss for the training and thus can obtain lower values for the metrics that are related to distortion of the spectrogram. This results suggest that MR-STFT and MCD may be insufficient to evaluate the perceptual quality of a singing voice. 
We also evaluate the effect of increasing the hierarchy to 3-stage. HPG-3 further improves the objective metrics with a 30\% increase in computational cost. For a fair comparison, we evaluate the larger PriorGrad model (PriorGrad-L) by increasing the number of layers and channels to $L=40$ and $80$, respectively.  However, the performance of PriorGrad-L does not change significantly. This suggests that the proposed HPG more efficiently scales to the larger model.

\begin{figure}[t]
  \centering
  \includegraphics[width=\linewidth]{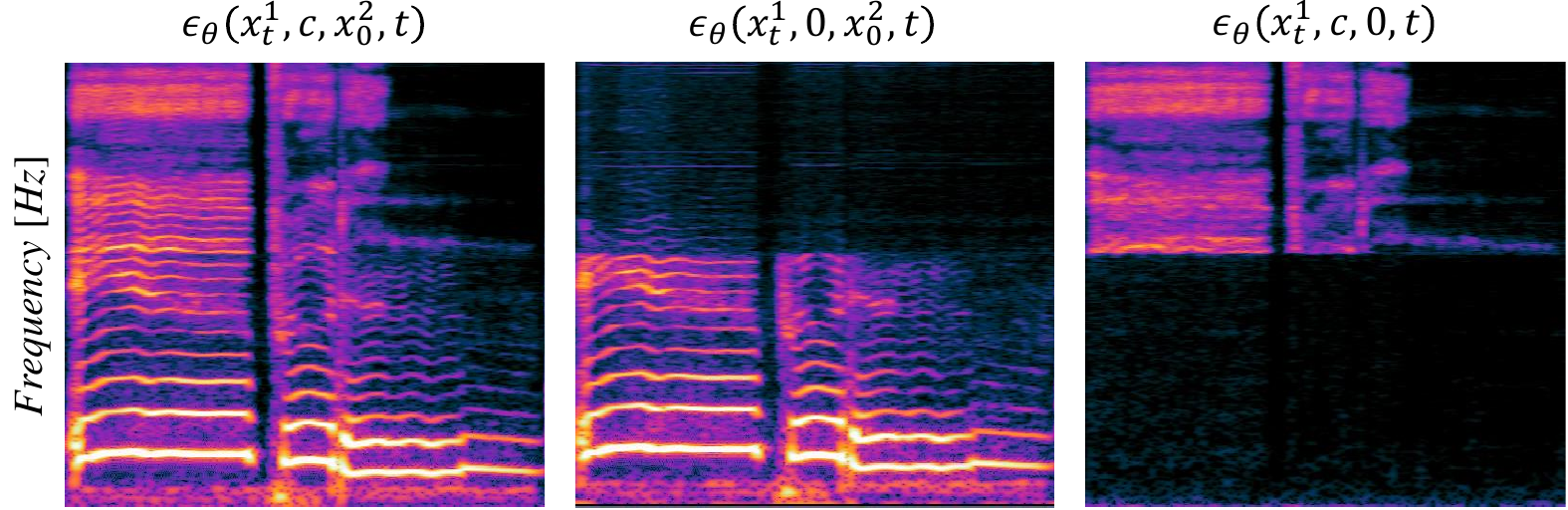}
  \caption{Spectrograms of generated data with different conditioning.}
  \label{fig:condition}
\end{figure}

Finally, we investigate how the HPG model uses the conditioning data. We replace either the mel-spectrogram $c$ or the data at the lower sampling rate $x_0^2$ of the HPG-2 model to 0 and generate the samples. 
As shown in Figure \ref{fig:condition}, the model generates signal under the Nyquist frequency of the lower module $\frac{f_s^2}{2}$ even when the mel-spectrogram is replaced to zero. On the other hand, when the low sampling rate data $x_0^2$ is replaced to zero, the model generates only high frequency components. This results show that the model utilizes information from $x_0^2$ to generate the low-frequency components while the high-frequency ones are generated on the basis of $c$ as expected. 
Audio samples are available at our website$^{\ref{fn:demo}}$.

\begin{table}[t]
   \caption{\label{tab:obj} {\it Objective test results. For all metrics, lower the better.}}
    \vspace{2mm}
    \centering{
      \footnotesize
    \begin{tabular}{c | c | c c c c c} 
    \hline
    Model	&RTF &PMAE	&VDE &MR-STFT	&MCD\\
    \hline
    PWG \cite{Yamamoto20PWG}	&0.067	&3.12	&5.61	&1.09	&\textbf{6.63}\\
    PriorGrad \cite{Lee22PriorGrad}	&\textbf{0.066}	&1.80	&3.96	&1.34	&9.62\\
    PriorGrad-L	&0.093	&2.08	&3.86	&1.38	&9.47\\
    \hline
    HPG-2 (Ours)	&0.070	&1.82	&3.47	&1.13	&8.97\\
    HPG-3 (Ours)	&0.100	&\textbf{1.67}	&\textbf{3.32}	&\textbf{1.07}	&8.12\\
    \hline
    \end{tabular}
    } 
\end{table}

\section{Conclusion}
We proposed a hierarchical diffusion model for singing voice neural vocoders. The proposed method learns diffusion models in different sampling rates independently while conditioning the model with data at the lower sampling rate. During the inference, the model progressively generates a signal while taking care of the anti-aliasing filter. Our experimental results show that the proposed method applied to PriorGrad outperforms PriorGrad and Parallel WaveGAN at similar computational costs. Although we focus on singing voices in this work, the proposed method is applicable to any type of audio. Evaluating the proposed method on different types of audio such as speech, music, and environmental sounds will be our future work.

\ninept

\bibliographystyle{IEEEbib}
\bibliography{vc,bss,other}


\end{document}